\documentclass[letterpaper]{article} 
\usepackage{aaai2027}  
\usepackage[hyphens]{url}  
\usepackage{graphicx} 
\usepackage{natbib}  
\usepackage{caption} 
\usepackage{url}
\usepackage[most]{tcolorbox}
\usepackage{subcaption}
\usepackage{algorithm}
\usepackage{algorithmic}
\usepackage{amsmath}
\usepackage{multirow}
\usepackage{amssymb}
\usepackage{newfloat}
\usepackage{listings}
\DeclareCaptionStyle{ruled}{labelfont=normalfont,labelsep=colon,strut=off} 
\floatstyle{ruled}
\newfloat{listing}{tb}{lst}{}
\floatname{listing}{Listing}

\usepackage{booktabs}

\begin{document}

\nocopyright  

\title{Never Stop Speaking: a Denial-of-Service Attack on End-to-End Speech Language Models}

\author{
Shuozhe Cheng\textsuperscript{1}\thanks{Email: sc8483629@gmail.com},
Kunlan Xiang,
Mingxuan Li,
Ji Zhang,
Dongxiao Liu,
Wenbo Jiang\thanks{Corresponding author ,  Email:wenbo\_jiang@uestc.edu.cn}
}

\affiliations{}
\maketitle

\begin{abstract}
Many studies have shown that specially crafted inputs can induce large language models (LLMs) to generate excessively long outputs, resulting in significant computational overhead and resource consumption. While most existing denial-of-service (DoS) attacks target text-only LLMs, end-to-end (E2E) speech LLMs are rapidly emerging. Existing text-based DoS attacks primarily rely on prompt engineering, such as adversarial suffixes or semantic inducement, which exploit the discrete nature of text inputs and therefore cannot be directly transferred to continuous speech inputs. Moreover, prior studies on speech model security mainly focus on ASR or TTS systems, leaving the DoS vulnerability of E2E speech LLMs largely unexplored. To address this gap, we propose the perturbation-based DoS attack targeting E2E speech models. Instead of inducing long outputs through prompt manipulation, our method optimizes imperceptible acoustic perturbations to directly influence the model's autoregressive generation process while preserving the original input length. Specifically, we formulate the attack as a composite optimization objective that jointly suppresses EOS generation, encourages prolonged decoding, and largely preserves semantic consistency by integrating weighted EOS loss, top-k logit loss, length loss, and semantic alignment loss. To further improve stealthiness, we employ voice activity detection (VAD) to inject perturbations only into voiced regions. Extensive experiments on three open-source E2E speech LLMs demonstrate that our method achieves stable attack success rate while significantly increasing generation length and GPU resource consumption, revealing security risks in modern ALLMs.

\end{abstract}


\section{Introduction}

End-to-end Audio Large Language Models (E2E ALLMs) have recently emerged as a promising paradigm for human-machine interaction, enabling direct understanding and generation of speech without relying on explicit intermediate representations. With their rapid deployment in voice assistants and conversational systems, ensuring the security and reliability of these models has become increasingly important. Among various security threats, denial-of-service (DoS) attacks pose a critical risk by forcing models to generate excessively long responses, leading to unnecessary computational consumption and service degradation. However, while DoS attacks have been extensively studied in text-based LLMs, the vulnerability of E2E ALLMs remains largely unexplored.
\begin{figure}
  \centering
  \captionsetup{font=footnotesize}
  \includegraphics[width=0.48\textwidth,height=3.4cm]{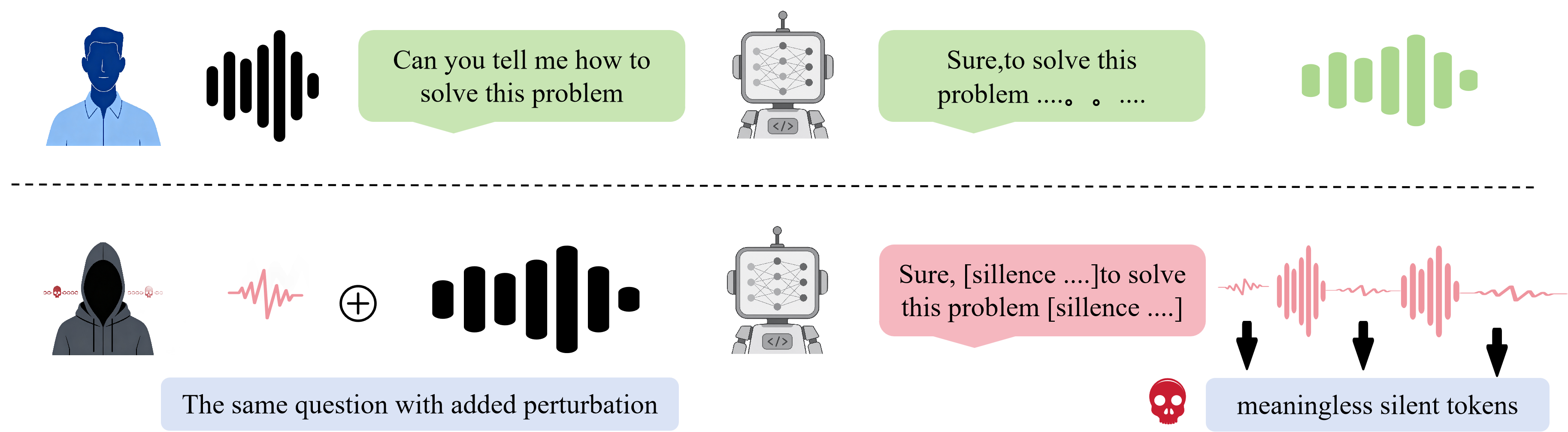}
  \caption{The figure shows the outputs of the model with and without added noise for the same input.A clean input x(top) yields a normal speech response, while the adversarial example x+\begin{math}\delta\end{math}(bottom), crafted with a tiny perturbation, induces the model to produce excessive meaningless silent segments, drastically prolonging autoregressive decoding and incurring heavy computational overhead.}
  \label{fig:1}
\end{figure}

Existing DoS attacks against large language models \cite{liu2026reasoningbomb,chen-etal-2026-naturalsloth,sim2025excessive} have been extensively studied. For example, adversarial suffix-based methods manipulate model behavior by appending carefully designed tokens, while recent approaches construct complex queries or decompose instructions into fine-grained reasoning steps to induce unnecessarily long generation. However, these strategies cannot be directly applied to end-to-end audio large language models (E2E ALLMs). Unlike text models that operate on editable discrete tokens, E2E ALLMs process continuous acoustic waveforms, making token-level manipulation infeasible. Moreover, prompt-based strategies typically rely on abnormally complex or unnatural inputs, which increase attack detectability and are difficult to transfer to speech inputs.Existing studies on ASR models \cite{haque2023slothspeech} have explored DoS attacks by suppressing the logits of the EOS token. However, these methods mainly target the speech-to-text component and cannot address the complex generation process of E2E ALLMs. Furthermore, optimizing only EOS logits may lack sufficient constraints on other aspects of generation, potentially degrading semantic consistency or causing unstable decoding behavior. Therefore, crafting effective DoS attacks against E2E ALLMs remains challenging for three reasons. First, perturbations must be optimized in the continuous waveform space while remaining imperceptible and preserving the original input length. Second, directly optimizing discrete generation length is non-differentiable. Third, excessive suppression of termination signals may disrupt semantic consistency or generation stability, resulting in premature decoding failures rather than prolonged outputs.

To address these issues, we propose an improved method based on multi-loss EOS logit suppression. Our loss function consists of the following four components:  (1) Weighted EOS logit loss: On top of the basic summation of all EOS logits, we assign dynamic weights to the logits at each autoregressive step, incorporating token-wise penalties and generation process penalties.
(2)  top-k logit loss: We indirectly suppress the EOS output by increasing the probability of the top-k tokens output by the model.  
(3)  Length loss: We force the model’s output length to align with max new tokens.  
(4) Semantic similarity loss: We maintain the semantic consistency between the original audio and the adversarial audio to ensure attack stealthiness. We show an example in Figure \ref{fig:1}.

To enhance attack stealthiness, we further employ voice activity detection (VAD) to extract voiced segments of the audio and add noise only to these voiced portions. Experimental results demonstrate that our method achieves strong performance on ALLM (Audio Large Language Model). Our contributions are as follows:  
\begin{itemize}
\setlength{\itemsep}{1pt}

\item We present a novel end-to-end Denial-of-Service  attacks on ALLMs, filling a critical gap in end-to-end ALLM security.
\item We propose a multi-objective optimization loss function to instantiate these attacks, providing a principled method for crafting adversarial inputs that disrupt ALLM behavior.
\item Extensive experiments demonstrate the effectiveness of our attacks, opening new research directions and highlighting practical vulnerabilities in ALLM systems.

\end{itemize}

\section{Related Work}
\label{gen_inst}
\subsection{Denial-of-service attack}

The goal of Denial of Service (DoS) attacks is typically to break the output constraint of a model by constructing specially crafted inputs—such as adding adversarial suffixes, introducing spelling errors, or using homophones—thereby forcing the model to generate output indefinitely. As we know, each output token consumes substantial computational resources. Once a model begins generating unbounded responses, it continuously consumes server resources on the one hand; on the other hand, server resources are finite—once depleted, the server can no longer serve legitimate users.  Previous research has demonstrated the effectiveness of DoS attacks. For example, Zhang et al. proposed the AutoDoS\cite{zhang2024crabs} algorithm, which constructs a DoS attack tree and expands node coverage to achieve effectiveness under black-box settings. The P-DoS\cite{gao2024dospoisoning} attack injects poisoned samples during model fine-tuning to reduce the logits of the EOS token, thereby breaking the model's output length limit. However, due to the discrete nature of text-based methods, they cannot be directly transferred to the audio domain. Existing attacks on audio models use overly simplistic loss functions and have not achieved satisfactory performance.
\subsection{E2E ALLM}
The development of the Transformer has brought transformative changes to the field of artificial intelligence. With its unique attention mechanism and contextual processing capability, it has rapidly driven the evolution of LLM architectures, demonstrating superiority across various domains. Leveraging the powerful capabilities of LLMs, many studies have explored multimodal \cite{latif2023sparks} fusion architectures for text–video, text–image, and text–speech. Benefiting from this, speech models have gradually evolved from discrete cascade architectures (where speech input must pass through a text-based intermediate medium to produce output) to end-to-end (E2E) architectures centered around the Transformer. E2E ALLMs (Audio Large Language Models) have no explicit speech-to-text process internally
\cite{dao2024ichigo}. Instead, they achieve bidirectional understanding of text and speech through text–speech alignment strategies, often employing interleaved text and speech generation strategies, demonstrating powerful speech processing capabilities.
\subsection{VAD}

Voice Activity Detection  aims to accurately distinguish voiced segments from silence or background noise in an audio signal, and is one of the fundamental techniques in speech processing. Traditional VAD methods are mostly based on energy thresholds, zero-crossing rates, or statistical models \cite{1162800,webrtcvad}, but their performance is limited under low signal-to-noise ratio conditions. With the development of deep learning, neural network based VAD methods \cite{9415081,zhang2016feature} have significantly improved detection robustness. In the field of adversarial attacks, VAD has been used to constrain the region where perturbations are added, for instance, adding noise only to voice activity regions to improve stealthiness while maintaining attack effectiveness. Our work adopts this idea by using VAD to extract voiced segments from the audio and adding optimized perturbations only on those segments, thereby avoiding noticeable artifacts in silent regions.
\begin{figure*}
  \centering
  \captionsetup{font=footnotesize }
  \includegraphics[width=0.93\textwidth,height=5.7cm]{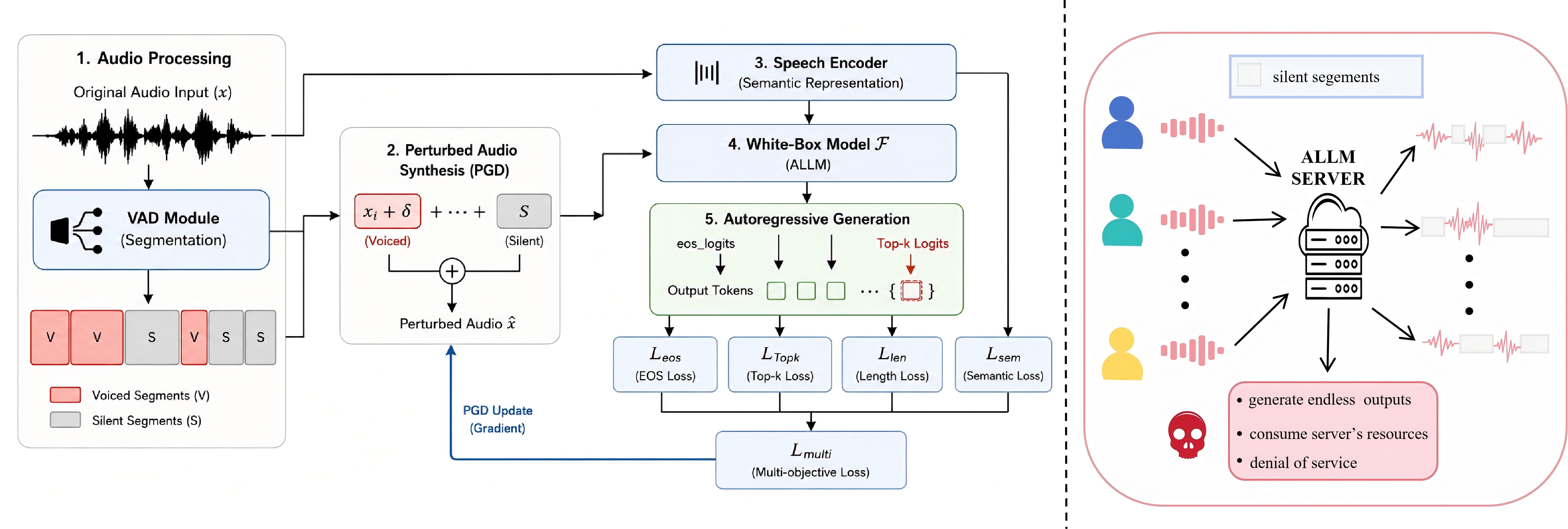}
  \caption{This figure illustrates the workflow of the adversarial DoS attack targeting end-to-end speech large models. VAD is used to separate voiced and silent segments and generate a mask for adding imperceptible perturbations. Adversarial examples are synthesized by applying tiny perturbations to the original speech. In the white-box optimization loop, we construct the multi-loss function \(\mathcal{L}_{multi}\) consisting of \(\mathcal{L}_{eos}\), \(\mathcal{L}_{Topk}\), \(\mathcal{L}_{len}\), and \(\mathcal{L}_{sem}\), and optimize perturbations via PGD. }
  \label{fig:2}
\end{figure*}
\section{Method}
\label{headings}
 In this work, we build on previous research and propose an improved method that constructs a composite loss function to achieve more precise and effective attacks. To enhance stealthiness, we employ VAD to segment the audio and add noise only to the voiced portions\cite{11319415}. Our optimization method uses gradient descent. 

\subsection{Threat Model}
We consider a denial-of-service (DoS) threat against end-to-end ALLMs. The attacker’s goal is to degrade service availability by inducing the model to generate excessively long responses containing meaningless silent segments, thereby exhausting computational resources and increasing inference costs.

We focus on the white-box attack setting, where the attacker has full access to the target model architecture and parameters and can compute gradients with respect to the input speech. However, the attacker can only introduce small, imperceptible adversarial perturbations to the input audio without modifying the model or its deployment environment. The perturbation magnitude is constrained below the human auditory perception threshold to ensure stealthiness.
\subsection{Attack Overview}
Prior work has shown that large language models (LLMs)\cite{openai2023periodic} are susceptible to denial-of-service (DoS) attacks. Conventional DoS methods targeting text-based models typically append adversarial suffixes to the input text—a technique that introduces discrete tokens and is thus inapplicable to the continuous audio modality. Alternative approaches craft complex semantic constructs or sub-queries to elicit longer outputs; however, naively converting such crafted text into speech and feeding it to the model lacks novelty, and our subsequent experiments confirm that this strategy is largely ineffective.

Therefore, we propose our method. We consider a white-box attack scenario, where the attacker has full access to the architecture and parameters of the target end-to-end audio large language model (E2E ALLM) and can compute the gradients of the model's output. Given an initial audio x, a white-box model F to be attacked, and a composite loss function L, we first apply VAD to segment x into voiced and silent segments. We extract all voiced segments \begin{math} V=\{x_1, x_2, \dots, x_n\}\end{math},and silent segments \begin{math} S=\{y_1, y_2, \dots, y_n\}\end{math}. The goal of our attack is to generate noise  \begin{math} N=\{\sigma_1, \sigma_2, \dots, \sigma_n\}\end{math} for each voiced segment such that adding these noises to the original speech forces the model F to  output as long as possible. It can be formulated as a constrained optimization problem:
\begin{align}
\min_{\{\sigma_1, \sigma_2, \dots, \sigma_n\}} \quad \mathcal{L}\left(F(\hat{x})\right) ,\\
{s.t.} \quad  \|\sigma_i\|_p \leq \epsilon_i, \quad \forall i \in \{1,2,\dots,n\},\\
\hat{x} = {Concat}\left( \left\{x_i+\sigma_i\right\}_{i=1}^n,\ S\right)
\end{align}
where \(\mathcal{L}\) is our proposed composite loss function detailed in the next subsection. By minimizing \(\mathcal{L}\), the optimized perturbation effectively suppresses the End-of-Sentence (EOS) token probability while preserving semantic coherence, leading to prolonged autoregressive generation. Figure \ref{fig:2} illustrates the overall workflow of our attack.

\subsection{Noise Optimization Based on Composite Loss Function}
\textbf{Optimization Pipeline} Given the initial audio \(\mathbf{x}\) and the target model \(F\), we iteratively optimize the perturbation via gradient descent. At the beginning of each iteration, we concatenate the perturbed voiced segments (initialized with random Gaussian noise) with the unaltered silent segments to reconstruct the full adversarial waveform \(\hat{x}\). We then feed \(\hat{x}\) into the speech encoder of the target model to extract acoustic features, followed by the autoregressive decoder. During the decoding process, we record three types of outputs at each step: the logit of the EOS token, the logits of the top-\(K\) tokens, and the final total generation length \(N\). After the generation finishes (either by emitting EOS or reaching the maximum limit), we compute the composite loss \(\mathcal{L}_{DoS}\) using the recorded values. The gradient of \(\mathcal{L}_{DoS}\) with respect to the perturbation is calculated via backpropagation, and the perturbation is updated using Projected Gradient Descent (PGD)\cite{madry2018towards}. This optimization loop repeats until the generated response reaches the preset maximum token limit \(N_{\max}\) or the maximum iteration count is exhausted.

The overall composite loss is defined as:
\begin{equation}
\mathcal{L}_{DoS} = \mathcal{L}_{eos} + \mathcal{L}_{topk} + \mathcal{L}_{len}  + \mathcal{L}_{sem}
\end{equation}

In this loss function, the terms can be mainly divided into direct losses and indirect losses. The direct losses focus on controlling the generation length of the model by suppressing the EOS logits and stretching the length of the speech tokens output by the model. The indirect losses achieve indirect suppression of the EOS logits and smooth the model behavior by increasing the probability of top‑k token outputs and preserving semantic invariance of the noisy speech, thereby raising the probabilities of other tokens and preventing premature termination caused by semantic issues.

Direct losses. The direct losses consist of two parts. The first is the EOS logits loss. Unlike the traditional approach of simply summing all EOS logits, we assign a weight to each EOS logit as follows:
\begin{equation}
\mathcal{L}_{eos} = \frac{1}{N} \sum_{t=1}^{N} w_t \ell_{EOS,t}
\end{equation}

The weight consists of a sign-based gating component and a dynamically growing component. The gating component depends on the sign of the logit value. Since a non-positive EOS logit implies that the EOS token is almost impossible to be output, we set its weight to zero to completely exclude such uninformative steps from the optimization. Only steps with positive EOS logits are assigned a positive weight, directing the optimizer toward positions where EOS suppression is actually meaningful. The dynamic component depends on the depth of the autoregressive step. It is known that the probability of outputting the EOS token increases as generation proceeds. Therefore, this component gradually increases with the autoregressive step, and we adopt exponential growth to make the optimization pay more attention to later stages of the model output. 
\[
    \quad
    w_{t}=
    \begin{cases}
    w_{high}\exp \left( \frac{t}{N}\kappa \right) ,&\ell _{EOS,t}>0\\
    0 ,&\ell _{EOS,t}\leq 0
    \end{cases}
    \]
    
To encourage long generation sequences, we introduce a differentiable length loss. 
Since the generated audio token length is determined by the EOS prediction during autoregressive decoding, directly optimizing the discrete output length is not feasible. 
Therefore, we estimate the expected generation length based on the probability distribution of EOS termination at each generation step. Let $z_t$ denote the EOS logit at the $t$-th generation step and $p_t=\sigma(z_t)$ be the corresponding EOS probability. 
The probability that the generation terminates at step $t$ is computed by considering that no EOS token is generated before step $t$ and an EOS token is generated at step $t$. 
The expected generation length is then formulated as:

\begin{equation}
\mathbb{E}[L]
=
\sum_{t=1}^{N}t
\left(
p_t\prod_{i=1}^{t-1}(1-p_i)
\right)
\end{equation}

where $N$ is the maximum generation length. 
The length loss is defined as:

\begin{equation}
\mathcal{L}_{Len}
=
\frac{
(N_{\max}-\mathbb{E}[L])^2
}
{N_{\max}} .
\end{equation}

This formulation encourages the expected generation length to approach the maximum allowed length while maintaining a comparable scale with other loss terms.

Indirect losses. The indirect losses also consist of two parts. The first is the top‑k loss. At each generation step, the model samples the output from the top‑k tokens. Therefore, the top‑k loss focuses on increasing the logits of the top‑k tokens. The increase in top‑k logits indirectly reduces the EOS logits. At the same time, increasing the top‑k logits helps preserve the coherence of the synthesized audio.

\begin{equation}\mathcal{L}_{topk} = -\frac{1}{N}\sum_{t=1}^{N}\frac{1}{K}\sum_{k=1}^{K} \ell_{topk,t}^{(k)}
\end{equation}

Finally, we introduce a semantic alignment loss. For the original speech and the noise‑added speech, we extract their features using a speech encoder and compute the cosine similarity between the two feature vectors. This ensures that the perturbation does not affect the semantics of the audio, thereby eliminating the possibility that changes in output length are caused by semantic alteration.
\begin{equation}
\mathcal{L}_{sem} = 1 - \frac{\langle s(x), s(x+\delta) \rangle}{\|s(x)\|_2 \|s(x+\delta)\|_2}
\end{equation}

\section{Experiment}

\subsection{Experiment Setup}
\paragraph{Model and Datasets}
We evaluate our audio adversarial attack on three recent open‑source audio large models: LFM2.5‑Audio‑1.5B\cite{liquidai2025lfm2},Fun‑Audio‑Chat‑8B\cite{tan2025drvoiceparallelspeechtextvoice} and Qwen2‑Audio‑7B‑Instruct\cite{Qwen2-Audio}. LFM2.5‑Audio is a lightweight end‑to‑end foundation model designed for low‑latency real‑time speech conversation; Fun‑Audio‑Chat and Qwen2‑Audio are medium‑sized audio‑language dialogue models optimized for general speech interaction and audio analysis. We evaluate on two primary models and further report Qwen2-Audio results in Appendix. For evaluation, we use two public speech datasets, OpenSLR and QCRI.

\paragraph{Attack Setup}
During the attack phase, we adopt the Projected Gradient Descent (PGD) framework to optimize adversarial perturbations, aiming to induce the target model to generate longer outputs. Specifically, we randomly sample 100 clean audio samples from each dataset, add the adversarial perturbations (generated by our method) to the original audio, and feed the perturbed audio into the target models to measure the attack success rate. Due to limited computational resources, we set \(N_{\max} = 1024\).

The hyperparameters are set as follows. For \(\mathcal{L}_{eos}\), we set the 
\(w_{high} = 4\) and the exponential growth parameter \(\kappa = 4\). For \(\mathcal{L}_{topk}\), we set \(k = 3\). During optimization, we run PGD for 200 iterations per sample, with the perturbation magnitude constrained by an \(\ell_{\infty}\) norm bound of \(\epsilon = 10^{-4}\).
\begin{table*}[t]
\centering
\small
\caption{Attack performance comparison across different models and baselines. All experiments are performed on the OpenSLR and QCRI datasets across evaluated models.}
\setlength{\tabcolsep}{8pt}
\label{tab:main_results}
\resizebox{0.85\textwidth}{!}{%
\begin{tabular}{ccccc}
\toprule
\textbf{Model} & \textbf{Method} & \textbf{ASR ($\uparrow$)} & \textbf{AVG. Output Length ($\uparrow$)} & \textbf{Memory Usage (GB)} \\
\midrule
\multirow{5}{*}{Liquid Audio (1.5B)} 
& Clean audio        & 0.00 & 198.34 & 8.89/47.99 \\
& Random Noise       & 0.00 & 205.51 & 8.58/47.99 \\
& Simple Loss        & 0.79 & 857.38 & 10.56/47.99 \\
& Crabs  & 0.34 & 545.72 & 9.96/47.99 \\
& ExtendAttack  & 0.29 & 496.37 & 10.18/47.99 \\

& \textbf{Our method} & \textbf{0.87} & \textbf{941.88} & 10.78/47.99 \\
\midrule
\multirow{5}{*}{FunAudioChat (8B)} 
& Clean audio        & 0.00 & 213.57 & 20.26/47.99 \\
& Random Noise       & 0.00 & 210.83 & 20.29/47.99 \\
& Simple Loss        & 0.77 & 869.52 & 21.74/47.99 \\
& Crabs  & 0.36 & 594.38 & 21.50/47.99 \\
& ExtendAttack  & 0.48 & 607.54 & 21.16/47.99 \\

& \textbf{Our method} & \textbf{0.84} & \textbf{920.24} & 21.93/47.99 \\
\bottomrule
\end{tabular}
}
\end{table*}

\begin{table*}[h]
\centering
\caption{Ablation study on loss components \(\mathcal{L}_{sem}\), \(\mathcal{L}_{len}\) (length loss), \(\mathcal{L}_{eos}\), and \(\mathcal{L}_{topk}\).}
\label{tab:ablation_loss}
\resizebox{0.85\textwidth}{!}{
\begin{tabular}{ccccc|ccc}
\toprule
\(\mathcal{L}_{sem}\) & \(\mathcal{L}_{len}\) & \(\mathcal{L}_{eos}\) & \(\mathcal{L}_{topk}\) & VAD
& ASR (\(\uparrow\)) & Avg.Output Length (\(\uparrow\)) &  Response Quality (\(\uparrow\)) \\
\midrule
\(\checkmark\) & \(\checkmark\) & \(\checkmark\) &  & \(\checkmark\)& 0.80 & 893.47 & 4.46 \\
\(\checkmark\) & \(\checkmark\) &  & \(\checkmark\) & \(\checkmark\)& 0.12 & 289.43 & 4.71 \\
\(\checkmark\) &  & \(\checkmark\) & \(\checkmark\) & \(\checkmark\)& 0.79 & 867.65 & 4.70 \\
 & \(\checkmark\) & \(\checkmark\) & \(\checkmark\) & \(\checkmark\)& 0.83 & 937.81 & 3.92 \\
\(\checkmark\)& \(\checkmark\) & \(\checkmark\) & \(\checkmark\) & & 0.84 & 945.33 & 4.52 \\
\midrule
\(\checkmark\) & \(\checkmark\) & \(\checkmark\) & \(\checkmark\) & \(\checkmark\)& 0.84 & 950.24 & 4.75 \\
\bottomrule
\end{tabular}

}
\end{table*}
\paragraph{Metrics}
To comprehensively evaluate the effectiveness and resource cost of our attack method across different baselines and all reported results are averaged over 100 samples. We adopt the following three metrics:
\begin{itemize}

\item \textbf{Attack Success Rate (ASR):} The proportion of adversarial samples that successfully force the target model to generate audio token sequences reaching the maximum generation length limit.

\item \textbf{Output Token Length:} The number of tokens in the model’s generated response. We report the average output length over all test samples, which helps analyze how the attack influences the model’s generative behavior.

\item \textbf{Peak GPU Memory Usage:} The maximum GPU memory allocated during a single forward inference pass (without gradient computation), measured in MB or GB. This metric indicates the practical computational overhead and feasibility of the attack, especially for edge-side audio models.

\item 
\textbf{Response Quality:}
We extract the generated text tokens from clean and adversarial inputs and use ChatGPT 5.5\cite{openai2025gpt55} as an evaluator to assess semantic consistency and response quality. The final score is averaged over all samples on a 1-5 scale. The evaluation prompt is provided in Appendix~\ref{app:response_quality}.
\end{itemize}

To demonstrate the effectiveness of our proposed attack method, we compare it with the following three baselines: 
(1) using clean original audio or audio with random noise as input;
(2) a conventional method that relies solely on naive eos logits;
(3) a noise injection approach that operates on the entire global audio without employing Voice Activity Detection (VAD).
(4) Crabs\cite{zhang2024crabs}, a kind of DoS attack on natural LLM
(5) ExtendAttack\cite{zhu2025extendattackattackingserverslrms},a kind of dos attack on  LRM

\subsection{Main Result}
Table 1 summarizes the attack performance under the default top-k sampling strategy. 
Our method consistently achieves the strongest DoS capability across different E2E ALLMs, demonstrating its effectiveness in manipulating autoregressive generation behaviors. 
On LFM2.5-Audio and FunAudioChat, our method achieves attack success rates of 87\% and 84\%, respectively, significantly outperforming all baseline methods. 
Meanwhile, the generated output length is extended to 941.88 and 920.24 tokens, which is more than four times longer than clean inputs and substantially exceeds random noise and existing DoS baselines. 
This indicates that the optimized acoustic perturbations can effectively prevent normal termination and force the model to continue decoding for an extended period, thereby introducing considerable computational overhead.

Compared with the simple EOS suppression baseline, our method achieves higher ASR and longer outputs on both models. 
This improvement demonstrates that directly optimizing EOS logits alone is insufficient for reliable DoS attacks, while jointly modeling termination suppression, generation length, token distribution, and semantic consistency provides a more effective optimization objective. 
In addition, text-based DoS attacks, including Crabs and ExtendAttack, exhibit limited effectiveness when transferred to the audio domain, achieving much lower ASR and shorter outputs. 
These results further verify that discrete token manipulation strategies designed for text LLMs cannot be directly applied to continuous speech inputs.

Beyond generation length, our attack also introduces practical computational costs. 
As shown in Table 1, the increased output length leads to substantial GPU memory consumption during inference, with our method requiring 10.78 GB and 21.93 GB on LFM2.5-Audio and FunAudioChat, respectively, compared with only 8.89 GB and 17.26 GB for clean inputs. 
These results reveal that carefully optimized acoustic perturbations can effectively exploit the autoregressive generation mechanism of E2E ALLMs and cause significant resource consumption while preserving the original input semantics.

To further evaluate the robustness of our method, we additionally test different decoding strategies and the Qwen2-Audio model. We also test the attack transferability. The results under greedy decoding, on Qwen2-Audio and the transferability are reported in the Appendix.
\begin{figure}[t]
  \centering
  
  \captionsetup{font=footnotesize }
  \includegraphics[width=0.5\textwidth,height=3.5cm]{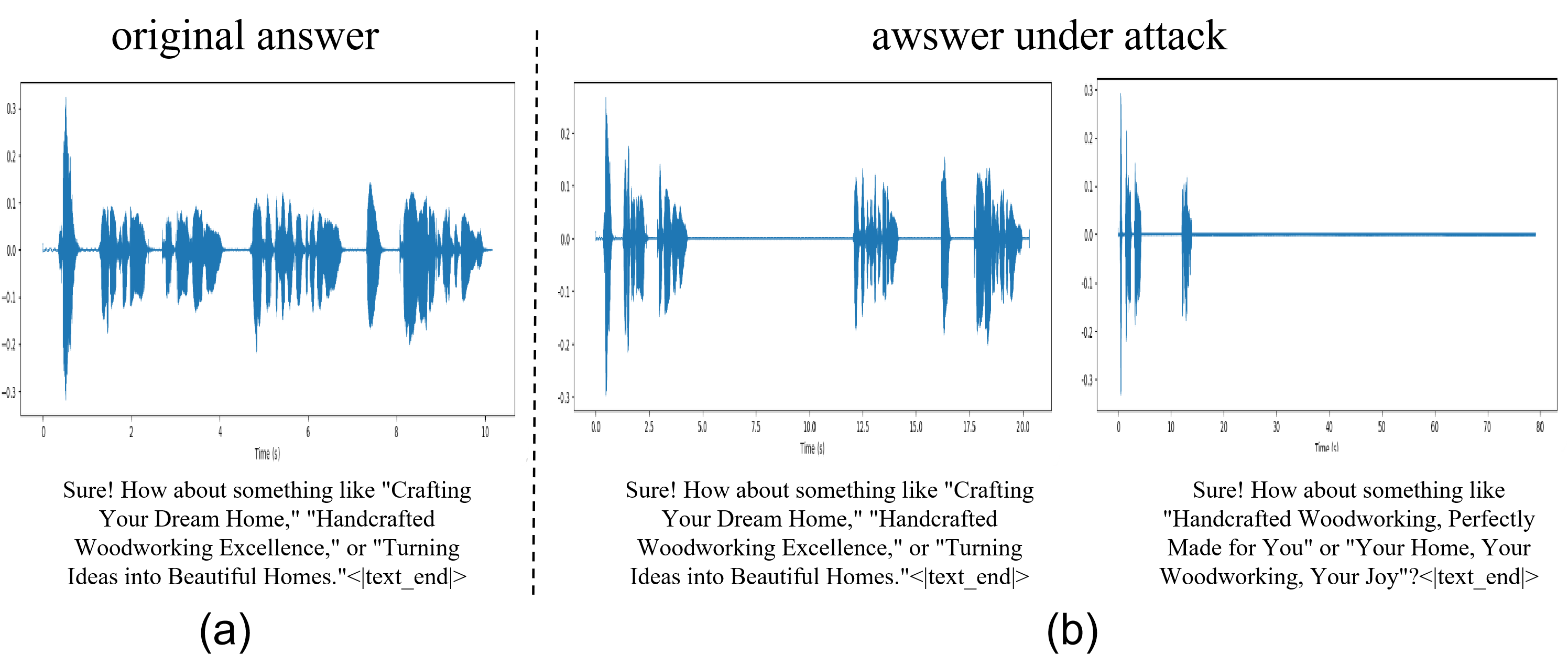}
  \caption{Waveform comparison between (a) clean output and (b) attacked output. The clean audio exhibits continuous voiced segments with high-amplitude oscillations, whereas the attacked output shows sparsely distributed speech fragments separated by extended flat silent regions }
  \label{fig:3}
\end{figure}
\subsection{Output Audio Analysis}
To further understand the mechanism behind output elongation, we analyze
the generated waveforms and intermediate text responses from successfully
attacked samples, as shown in Figure~3. The transcribed responses remain
largely consistent with those generated from clean inputs, indicating that
our attack does not significantly affect the model's semantic
understanding capability. However, waveform analysis reveals that attacked
outputs contain substantially more low-energy silent regions compared with
clean outputs. This phenomenon is mainly caused by the joint optimization
of $L_{EOS}$ and $L_{len}$, which encourages the model to continue
autoregressive decoding beyond its normal termination point. As generation
extends beyond meaningful response ranges, the additional tokens tend to
correspond to low-information speech segments, resulting in prolonged
outputs and increased computational consumption.

\subsection{Ablation Study}
\label{sec:ablation_loss_vad}
\textbf{Effectiveness of Loss Components.}
To investigate the contribution of each component in our composite
objective, we conduct ablation experiments by removing each loss term
from $L_{DoS}$ individually while keeping all other settings unchanged.
The results are summarized in Table \ref{tab:ablation_loss}.

Removing $L_{EOS}$ leads to a significant degradation in attack
performance, reducing the ASR from 0.84 to 0.12 and the average output
length from 950.24 to 289.43 tokens. This demonstrates that weighted EOS
suppression is the primary factor driving prolonged generation, as it
directly prevents premature termination during autoregressive decoding.
Without $L_{len}$, the ASR decreases slightly to 0.79 and the output
length drops to 867.65 tokens, indicating that the length loss provides
additional optimization guidance toward the maximum generation boundary.
Similarly, removing $L_{topk}$ causes a moderate reduction in ASR and
output length, while noticeably degrading response quality. This suggests
that $L_{topk}$ mainly improves generation stability by redistributing
token probabilities and indirectly suppressing EOS prediction.

Removing $L_{sem}$ slightly improves ASR and output length, but results
in a substantial decline in response quality (from 4.75 to 3.92). This
indicates that semantic alignment is not essential for maximizing output
length, but plays a critical role in preserving the original speech
semantics and ensuring attack stealthiness. Overall, the ablation results
verify that each component serves a complementary purpose: $L_{EOS}$
provides the fundamental attack capability, $L_{len}$ and $L_{topk}$
enhance generation extension and stability, while $L_{sem}$ maintains
semantic consistency.

\textbf{Effectiveness of the VAD Strategy.} 
To evaluate the contribution of the VAD-based perturbation strategy, we compare our full method with a variant that applies perturbations to the entire waveform without VAD. As shown in Table\ref{sec:ablation_loss_vad}, removing VAD results in a similar attack success rate (0.84) and slightly shorter output length, indicating that restricting perturbations to voiced regions does not compromise the attack effectiveness. Meanwhile, the response quality decreases from 4.75 to 4.52 without VAD, suggesting that perturbing silent regions introduces unnecessary acoustic interference and affects semantic preservation. These results demonstrate that VAD provides a better trade-off between attack effectiveness and stealthiness by concentrating perturbations on informative speech regions.

\textbf{Effect of the exponential growth parameter $\kappa$.}
We further investigate the impact of the exponential growth parameter $\kappa$ in $L_{eos}$.
As shown in Table~\ref{tab:kappa_ablation}, the attack performance first improves and then decreases as $\kappa$ increases, with $\kappa=4$ achieves the best or comparable performance across all evaluated models. When $\kappa$ is relatively small, the dynamic EOS weighting grows slowly, limiting the emphasis on later autoregressive steps where EOS prediction becomes more critical. Consequently, the optimizer may fail to sufficiently suppress EOS probabilities within the fixed optimization budget. In contrast, overly large $\kappa$ values cause the EOS weights to increase too rapidly, which may introduce excessively large gradients and make the optimization unstable. Moreover, an excessively dominant $L_{eos}$ term can weaken the contributions of other loss components, such as length guidance and semantic preservation, resulting in degraded attack effectiveness. Therefore, we select $\kappa=4$ as it provides a balanced trade-off between effective EOS suppression and stable multi-objective optimization.

\begin{table}[t]
\centering
\caption{Ablation study on the exponential growth parameter $\kappa$ in $L_{eos}$.}
\label{tab:kappa_ablation}
\begin{tabular}{c|ccccc}
\toprule
Model & $\kappa=2$ & $\kappa=3$ & $\kappa=4$ & $\kappa=5$ & $\kappa=6$ \\
\midrule
LFM2.5-Audio & 82\% & 84\% & \textbf{87\%} & 86\% & 79\% \\
FunAudioChat & 80\% & 82\% & \textbf{84\%} & 82\% & 76\% \\
Qwen2-Audio & 81\% & 83\% & \textbf{83\%} & 79\% & 77\% \\
\bottomrule
\end{tabular}
\end{table}
\subsection{Attack Robustness}

We further evaluate the robustness of our adversarial perturbations against additional random noise. 
Specifically, after generating adversarial examples, we add extra noise $\eta$ with different magnitudes relative to the original perturbation constraint $\epsilon$, i.e., $\|\eta\|_\infty=\alpha\epsilon$, and measure the resulting ASR.

As shown in Fig. 4(a), our attack remains effective under small additional perturbations, indicating that the optimized perturbations can tolerate moderate input variations. 
It is worth noting that $\alpha\epsilon$ represents the noise scale relative to the predefined perturbation bound rather than the actual magnitude of the optimized perturbation. 
Since the final perturbations generated by PGD may occupy different magnitude ranges across different models, the point at which additional noise starts to disrupt the attack varies among models. 
When the added noise reaches a comparable magnitude level to the optimized perturbation, the attack effectiveness gradually decreases, as the original adversarial direction is disturbed.

\subsection{Limitation}

Our robustness analysis reveals that lossy compression can affect the effectiveness of adversarial perturbations, as shown in Fig 4(b). This phenomenon is mainly because compression algorithms\cite{brand1995mp3,valin2012opus}  transform the original waveform representation by removing or quantizing fine-grained acoustic details, which may distort the optimized perturbation patterns. Similar effects can also be observed in other signal-level transformations, such as resampling, filtering, or noise injection, which modify the input distribution and consequently interfere with the adversarial directions.

However, these transformations introduce additional processing overhead and may degrade speech quality. Moreover, adaptive attackers may optimize perturbations considering the preprocessing pipeline, potentially reducing the robustness of such defenses. Therefore, developing practical defense mechanisms that can effectively disrupt adversarial perturbations while maintaining low latency and preserving speech quality remains an important future direction. 
\begin{figure}[t]
\centering
\begin{subfigure}{0.37\textwidth}
\centering
\includegraphics[width=\linewidth,height=4.7cm]{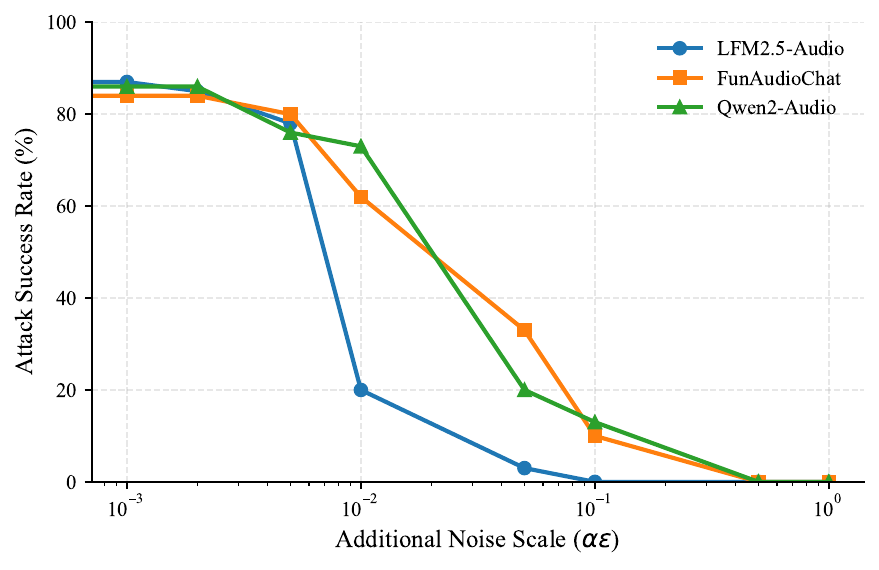}
\caption{ASR under additional noise}
\label{fig:sub_snr}
\end{subfigure}
\hfill
\begin{subfigure}{0.37\textwidth}
\centering
\includegraphics[width=\linewidth,height=4.7cm]{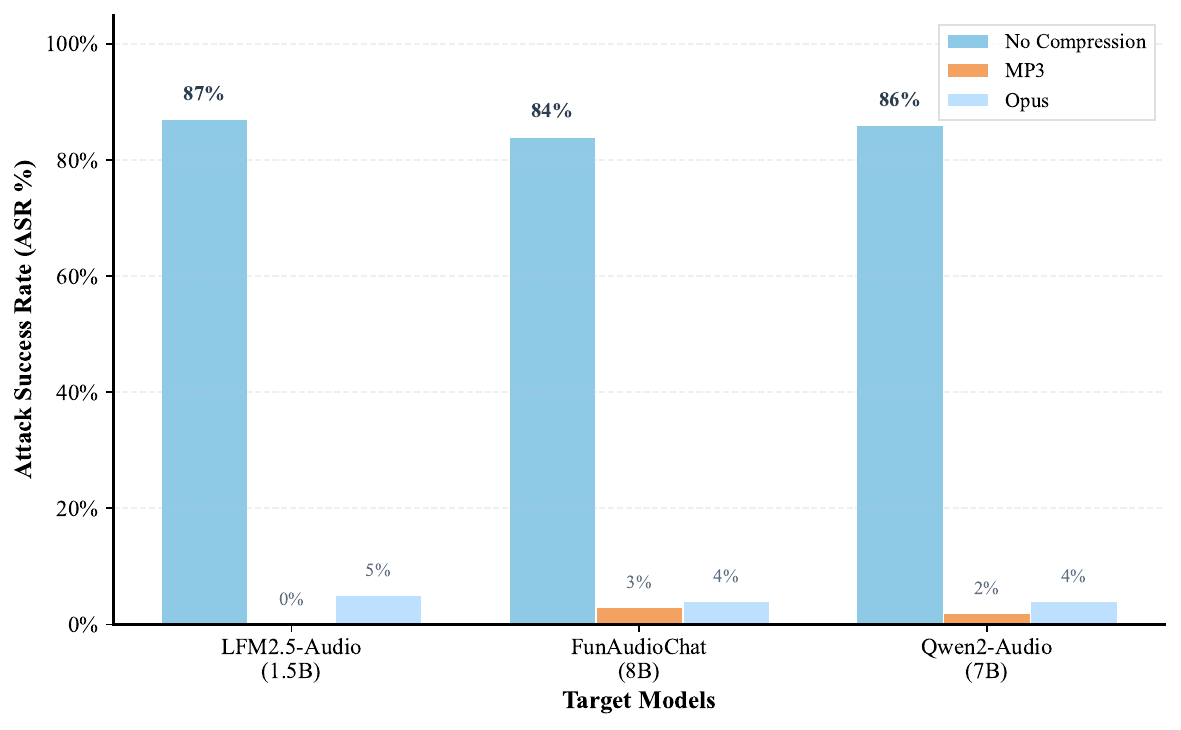}
\caption{Compression robustness}
\label{fig:sub_compress}
\end{subfigure}

\caption{(a) Attack performance under different additional noise. (b) Attack robustness against MP3 compression.}
\label{fig:transfer}
\end{figure}
\section{Conclusion}
\label{sec:conclusion}

In this paper, we present a systematic study of denial-of-service (DoS) attacks against end-to-end audio large language models (E2E ALLMs). We propose a white-box adversarial attack framework that leverages VAD-based perturbation and a multi-component optimization objective to suppress EOS prediction and induce prolonged autoregressive generation. Extensive experiments on multiple E2E ALLMs demonstrate that the proposed method can effectively increase output lengths, achieve high attack success rates, and introduce substantial inference overhead while largely preserving the semantic information of the original speech. 

Through comprehensive ablation studies, we analyze the contribution of each loss component and show that weighted EOS suppression, length guidance, token distribution optimization, and semantic alignment jointly improve attack effectiveness and stealthiness. Furthermore, our robustness analysis reveals that the attack can tolerate moderate input perturbations, while signal transformations such as compression may significantly reduce its effectiveness. These findings highlight the potential security risks of current E2E ALLMs and motivate the development of practical defense mechanisms against audio-domain adversarial DoS attacks.

\clearpage
\bibliography{aaai2027}

@misc{tan2025drvoiceparallelspeechtextvoice,
  title={DrVoice: Parallel Speech-Text Voice Conversation Model via Dual-Resolution Speech Representations}, 
  author={Chao-Hong Tan and Qian Chen and Wen Wang and Chong Deng and Qinglin Zhang and Luyao Cheng and Hai Yu and Xin Zhang and Xiang Lv and Tianyu Zhao and Chong Zhang and Yukun Ma and Yafeng Chen and Hui Wang and Jiaqing Liu and Xiangang Li and Jieping Ye},
  year={2025},
  eprint={2506.09349},
  archivePrefix={arXiv},
  primaryClass={cs.CL},
  url={https://arxiv.org/abs/2506.09349}, 
}

@article{liquidai2025lfm2,
 title={LFM2 Technical Report},
 author={Liquid AI},
 journal={arXiv preprint arXiv:2511.23404},
 year={2025},
}

@article{Qwen2-Audio,
  title={Qwen2-Audio Technical Report},
  author={Chu, Yunfei and Xu, Jin and Yang, Qian and Wei, Haojie and Wei, Xipin and Guo,  Zhifang and Leng, Yichong and Lv, Yuanjun and He, Jinzheng and Lin, Junyang and Zhou, Chang and Zhou, Jingren},
  journal={arXiv preprint arXiv:2407.10759},
  year={2024},
}

@article{zhang2024crabs,
  title={Crabs: Consuming Resource via Auto-generation for LLM-DoS Attack under Black-box Settings},
  author={Zhang, Yuanhe and Zhou, Zhenhong and Zhang, Wei and Wang, Xinyue and Jia, Xiaojun and Liu, Yang and Su, Sen},
  journal={arXiv preprint arXiv:2412.13879},
  year={2024},
  note={arXiv:2412.13879v4 [cs.CL]}
}

@misc{liu2026reasoningbomb,
  title={ReasoningBomb: A Stealthy Denial-of-Service Attack by Inducing Pathologically Long Reasoning in Large Reasoning Models},
  author={Xiaogeng Liu and Xinyan Wang and Yechao Zhang and Sanjay Kariyappa and Chong Xiang and Muhao Chen and G. Edward Suh and Chaowei Xiao},
  year={2026},
  eprint={2602.00154},
  archivePrefix={arXiv},
  primaryClass={cs.CR},
  doi={10.48550/arXiv.2602.00154}
}

@misc{gao2024dospoisoning,
  title={Denial-of-Service Poisoning Attacks against Large Language Models},
  author={Kuofeng Gao and Tianyu Pang and Chao Du and Yong Yang and Shu-Tao Xia and Min Lin},
  year={2024},
  eprint={2410.10760},
  archivePrefix={arXiv},
  primaryClass={cs.CR},
  doi={10.48550/arXiv.2410.10760}
}

@misc{webrtcvad,
  title={WebRTC Voice Activity Detector},
  year={2024},
  howpublished={\url{https://webrtc.org/}},
  note={Open-source real-time voice activity detection algorithm widely adopted in speech processing pipelines}
}

@INPROCEEDINGS{9415081,
  author={Wilkinson, Nicholas and Niesler, Thomas},
  booktitle={ICASSP 2021 - 2021 IEEE International Conference on Acoustics, Speech and Signal Processing (ICASSP)}, 
  title={A Hybrid CNN-BiLSTM Voice Activity Detector}, 
  year={2021},
  volume={},
  number={},
  pages={6803-6807},
  doi={10.1109/ICASSP39728.2021.9415081}}

@inproceedings{zhang2016feature,
  title={Feature Learning with Raw-Waveform CLDNNs for Voice Activity Detection},
  author={Zhang, Yu and Wang, Yuxuan and Glass, James},
  booktitle={Interspeech},
  pages={3542--3546},
  year={2016}
}

@ARTICLE{1162800,
  author={Atal, B. and Rabiner, L.},
  journal={IEEE Transactions on Acoustics, Speech, and Signal Processing}, 
  title={A pattern recognition approach to voiced-unvoiced-silence classification with applications to speech recognition}, 
  year={1976},
  volume={24},
  number={3},
  pages={201-212},
  doi={10.1109/TASSP.1976.1162800}}

@misc{latif2023sparks,
  title={Sparks of Large Audio Models: A Survey and Outlook},
  author={Siddique Latif and Moazzam Shoukat and Fahad Shamshad and Muhammad Usama and Yi Ren and Heriberto Cuay'ahuitl and Wenwu Wang and Xulong Zhang and Roberto Togneri and Erik Cambria and Björn W. Schuller},
  year={2023},
  eprint={2308.12792},
  archivePrefix={arXiv},
  primaryClass={cs.SD}
}

@misc{dao2024ichigo,
  title={Ichigo: Mixed-Modal Early-Fusion Realtime Voice Assistant},
  author={Alan Dao and Dinh Bach Vu and Huy Hoang Ha},
  year={2024},
  eprint={2410.15316},
  archivePrefix={arXiv},
  primaryClass={cs.CL}
}

@ARTICLE{11319415,
  author={Ko, Kyoungmin and Kim, Sunghwan and Kwon, Hyun},
  journal={IEEE Access}, 
  title={Audio Adversarial Example With No Noise in the Silent Area for Speech Recognition System}, 
  year={2026},
  volume={14},
  number={},
  pages={2924-2938},
  doi={10.1109/ACCESS.2025.3598087}}

@inproceedings{madry2018towards,
  title={Towards Deep Learning Models Resistant to Adversarial Attacks},
  author={Aleksander Madry and Aleksandar Makelov and Ludwig Schmidt and Dimitris Tsipras and Adrian Vladu},
  booktitle={International Conference on Learning Representations},
  year={2018},
  url={https://openreview.net/forum?id=rJzIBfZAb}
}

@misc{haque2023slothspeech,
  title={SlothSpeech: Denial-of-service Attack Against Speech Recognition Models},
  author={Mirazul Haque and Rutvij Shah and Simin Chen and Berrak \c{S}i\c{s}man and Cong Liu and Wei Yang},
  year={2023},
  eprint={2306.00794},
  archivePrefix={arXiv},
  primaryClass={cs.SD},
  doi={10.48550/arXiv.2306.00794}
}

@techreport{valin2012opus,
  title={Definition of the Opus Audio Codec},
  author={Valin, Jean-Marc and Vos, Koen and Terriberry, Timothy B.},
  institution={Internet Engineering Task Force},
  number={RFC 6716},
  year={2012}
}

@article{brand1995mp3,
  title={The MPEG audio layer-3 coding standard},
  author={Brandenburg, Karlheinz},
  journal={Proceedings of the IEEE},
  volume={83},
  number={4},
  pages={1260--1263},
  year={1995}
}

@misc{sim2025excessive,
      title={Excessive Reasoning Attack on Reasoning LLMs}, 
      author={Wai Man Si and Mingjie Li and Michael Backes and Yang Zhang},
      year={2025},
      eprint={2506.14374v1},
      archivePrefix={arXiv},
      primaryClass={cs.CR},
      url={https://arxiv.org/abs/2506.14374v1}
}

@misc{zhu2025extendattackattackingserverslrms,
      title={ExtendAttack: Attacking Servers of LRMs via Extending Reasoning},
      author={Zhenhao Zhu and Yue Liu and Zhiwei Xu and Yingwei Ma and Hongcheng Gao and Nuo Chen and Yanpei Guo and Wenjie Qu and Huiying Xu and Zifeng Kang and Xinzhong Zhu and Jiaheng Zhang},
      year={2025},
      eprint={2506.13737},
      archivePrefix={arXiv},
      primaryClass={cs.CR},
      url={https://arxiv.org/abs/2506.13737},
}

@inproceedings{chen-etal-2026-naturalsloth,
    title = "{N}atural{S}loth: Revisiting Denial-of-Service Attacks on Large Language Models",
    author = "Chen, Yiming  and
      Li, Zexin  and
      Yue, Xianghu  and
      Tan, Robby T.  and
      Li, Haizhou",
    editor = "Liakata, Maria  and
      Moreira, Viviane P.  and
      Zhang, Jiajun  and
      Jurgens, David",
    booktitle = "Proceedings of the 64th Annual Meeting of the {A}ssociation for {C}omputational {L}inguistics (Volume 1: Long Papers)",
    month = jul,
    year = "2026",
    address = "San Diego, California, United States",
    publisher = "Association for Computational Linguistics",
    url = "https://aclanthology.org/2026.acl-long.901/",
    doi = "10.18653/v1/2026.acl-long.901",
    pages = "19685--19702",
    ISBN = "979-8-89176-390-6"
}

@misc{openai2023periodic,
  author = {OpenAI},
  title  = {Periodic Outages Across {ChatGPT} and {API}},
  year   = {2023},
  note   = {Accessed on 16/03/2024},
  url    = {https://status.openai.com/incidents/21vl32gvx3hb}
}

@misc{openai2025gpt55,
  author       = {OpenAI},
  title        = {GPT-5.5 Technical Report},
  year         = {2025},
  howpublished = {OpenAI Technical Report},
  url          = {https://openai.com/}
}
\clearpage
\appendix
\section{Response Quality Evaluation Prompt}
\label{app:response_quality}
To evaluate response quality, we use an LLM-as-a-Judge strategy. 
The generated text responses from clean and adversarial inputs are provided to ChatGPT for evaluation. 
Each response is assigned a score from 1 to 5, and the final Response Quality score is calculated by averaging the scores over all test samples.

\begin{figure}[h]
\centering
\begin{minipage}{0.95\columnwidth}
\begin{tcolorbox}[
    colback=gray!10,
    colframe=gray!60,
    title={Prompt for Response Quality Evaluation},
    fonttitle=\bfseries,
    boxrule=0.5pt,
    arc=2mm,
    left=1mm,
    right=1mm,
    top=1mm,
    bottom=1mm
]
\small
\textbf{Instruction:} You are an evaluator for speech language model responses.

Given a reference response generated from clean audio and an adversarial response generated from perturbed audio, evaluate whether the adversarial response preserves the semantic information and task completion ability of the reference response.

Please assign an integer score from 1 to 5:

5: The adversarial response fully preserves the semantic content and quality of the reference response.

4: The adversarial response is mostly consistent with the reference response with minor degradation.

3: The adversarial response preserves the main information but has noticeable degradation.

2: The adversarial response has significant semantic differences or missing information.

1: The adversarial response fails to preserve the original meaning or task requirement.

Only output the score.
\end{tcolorbox}
\end{minipage}
\caption{Prompt used for LLM-based response quality evaluation.}
\label{fig:judge_prompt}
\end{figure}
\begin{table}[t]
\centering
\small
\caption{Attack performance under greedy decoding strategies}
\setlength{\tabcolsep}{8pt}
\label{tab:main_resultsA}
\resizebox{0.48\textwidth}{!}{%
\begin{tabular}{ccccc}
\toprule
\textbf{Model} & \textbf{Method} & \textbf{ASR ($\uparrow$)} & \textbf{AVG. Output Length ($\uparrow$)} & \textbf{Memory Usage (GB)} \\
\midrule
\multirow{5}{*}{Liquid Audio (1.5B)} 
& Clean audio        & 0.00 & 162.52 & 8.07/47.99 \\
& Random Noise       & 0.00 & 161.28 & 8.96/47.99 \\
& Simple Loss        & 0.77 & 832.58 & 10.46/47.99 \\
& Crabs  & 0.34 & 538.24 & 9.92/47.99 \\
& \textbf{Our method} & \textbf{0.86} & \textbf{927.69} & 10.27/47.99 \\
\midrule
\multirow{5}{*}{FunAudioChat (8B)} 
& Clean audio        & 0.00 & 187.22 & 20.53/47.99 \\
& Random Noise       & 0.00 & 195.79 & 20.76/47.99 \\
& Simple Loss        & 0.76 & 852.34 & 21.19/47.99 \\
& Crabs  & 0.38 & 613.72 & 21.07/47.99 \\
& \textbf{Our method} & \textbf{0.84} & \textbf{905.36} & 21.44/47.99 \\
\midrule
\multirow{5}{*}{QwenAudio(7B)} 
& Clean audio        & 0.00 & 182.35 & 18.37/47.99 \\
& Random Noise       & 0.00 & 181.43 & 18.12/47.99 \\
& Simple Loss        & 0.74 & 807.75 & 19.09/47.99 \\
& Crabs  & 0.43    & 643.35 & 20.07/47.99 \\
& \textbf{Our method} & \textbf{0.85} & \textbf{934.02} & 20.14/47.99 \\
\bottomrule
\end{tabular}
}
\end{table}
\section{Attack Performance under Greedy Decoding}

In the main experiments, we adopt top-k sampling as the default decoding strategy. To investigate whether the effectiveness of our attack depends on stochastic sampling behaviors, we further evaluate our method under greedy decoding, where the token with the highest probability is selected at each generation step.

As shown in Table 3, our method maintains strong attack performance under deterministic decoding. Specifically, it achieves ASRs of 86\%, 84\%, and 85\% on Liquid Audio, FunAudioChat, and Qwen2-Audio, respectively, while extending the average output length to over 900 tokens across all evaluated models. These results demonstrate that the proposed perturbations do not rely on sampling randomness, but instead directly influence the autoregressive generation process by suppressing termination and encouraging continuous decoding.

The consistent performance between top-k sampling and greedy decoding further verifies the robustness of our optimization objective, showing that the attack can generalize across different decoding strategies and effectively expose the inherent generation-control vulnerability of E2E ALLMs.
\begin{table}[t]
\centering
\small
\caption{Attack performance on Qwen2-Audio}
\setlength{\tabcolsep}{8pt}
\label{tab:main_resultsB}
\resizebox{0.48\textwidth}{!}{%
\begin{tabular}{ccccc}
\toprule
\textbf{Model} & \textbf{Method} & \textbf{ASR ($\uparrow$)} & \textbf{AVG. Output Length ($\uparrow$)} & \textbf{Memory Usage (GB)} \\
\midrule
\multirow{6}{*}{Qwen-Audio(7B)} 
& Clean audio        & 0.00 & 187.13 & 18.42/47.99 \\
& Random Noise       & 0.00 & 179.46 & 18.37/47.99 \\
& Simple Loss        & 0.80 & 846.53 & 20.01/47.99 \\
& Crabs  & 0.42 & 665.19 & 19.07/47.99 \\
& \textbf{Our method} & \textbf{0.83} & \textbf{913.07} & 19.94/47.99 \\
\bottomrule
\end{tabular}
}
\end{table}
\section{Evaluation under Qwen2-Audio}

To further evaluate the generality of our attack across different E2E ALLM architectures, we conduct additional experiments on Qwen2-Audio-7B-Instruct. As shown in Table 4, our method achieves an ASR of 83\% and increases the average output length to 913.07 tokens, significantly outperforming clean inputs and random noise baselines. Meanwhile, the generated responses consume 19.94 GB GPU memory during inference, demonstrating that the optimized perturbations can effectively induce prolonged decoding and introduce additional computational overhead.

Compared with the simple EOS suppression baseline, our method achieves higher attack effectiveness and longer generated sequences. This indicates that relying solely on EOS logit optimization is insufficient for stable DoS attacks, while jointly optimizing termination suppression, generation length, token distribution, and semantic consistency provides a more reliable attack objective. The consistent performance across Qwen2-Audio and the other evaluated E2E ALLMs demonstrates that our approach can effectively exploit the shared autoregressive generation mechanism of different audio language models.
\begin{figure}[t]
  \centering
  
  \captionsetup{font=footnotesize }
  \includegraphics[width=0.4\textwidth,height=5.5cm]{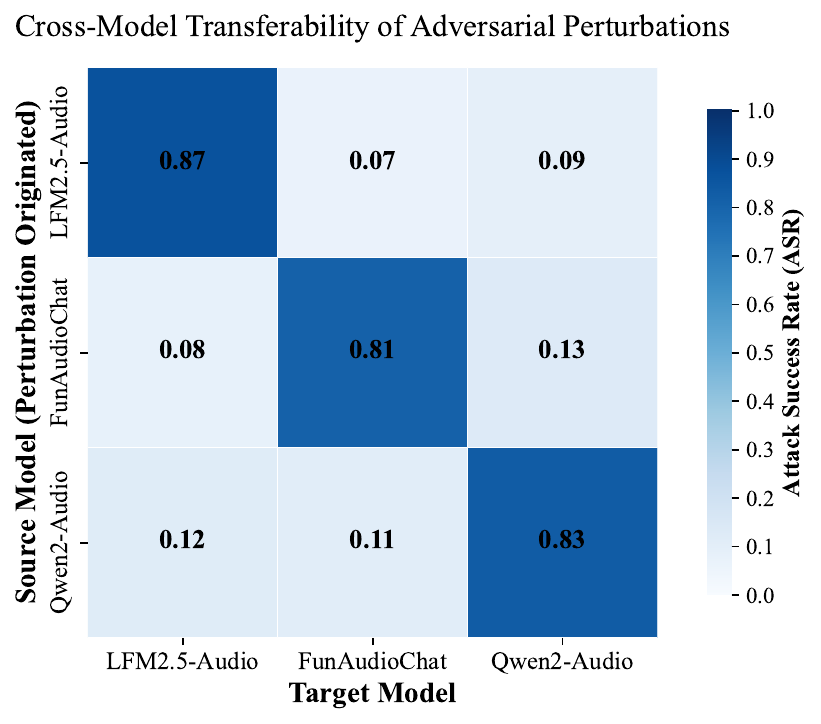}
  \caption{attack transferability }
  \label{fig:3}
\end{figure}
\subsection{Attack Transferability}

To evaluate the black-box transferability of our attack, we generate adversarial perturbations on one source model and directly apply them to other unseen target models. As shown in Fig\ref{fig:3}, the diagonal entries represent white-box attacks, achieving ASRs of 87\%, 81\%, and 83\% on LFM2.5-Audio, FunAudioChat, and Qwen2-Audio, respectively. For cross-model transfer, the perturbations achieve non-zero ASRs across all settings, with transfer rates ranging from 7\% to 13\%. Models with similar scales exhibit relatively higher transferability, indicating that the optimized perturbations capture partially shared generation characteristics among E2E audio models. These results demonstrate the feasibility of our method in practical black-box scenarios.
\end{document}